\documentstyle[prl,aps,twocolumn]{revtex}

\begin{document}
\iffalse
\draft

\title{ Induced Field Theory on the Brane World }

\author{		Keiichi Akama}
\address{	Department of Physics, Saitama Medical College,
		Kawakado, Moroyama, Saitama, 350-04, Japan}
\maketitle
\begin{abstract}
We show how the gravity, gauge, and matter fields are induced
on dynamically localized brane (domain wall, vortex etc.).
Besides the gravity and gauge fields, the extrinsic curvature fields
are induced, and they should obey the Gauss-Codazzi-Ricci equation
in addition to their own equations of motion. 
\end{abstract}

\vskip15pt
\else
\draft
\twocolumn[

\widetext
\hfill SMC-PHYS-162

\hfill hep-ph/0007001

\centerline{\large\bf	Induced Field Theory on the Brane World }
\vskip 10pt
\centerline{                Keiichi Akama }
\centerline{\small\sl       Department of Physics, Saitama Medical College,
                Kawakado, Moroyama, Saitama, 350-0496, Japan}
\centerline{\small       e-mail: akama@saitama-med.ac.jp}
\vskip 5pt

\leftskip 15mm\rightskip 15mm
\begin{abstract}
\baselineskip  = 10pt
We show how the gravity, gauge, and matter fields are induced on dynamically localized brane world (domain wall, vortex etc.). Solitonic solutions localized on a curved brane in certain field theories in higher dimensions are given in terms of gravity, gauge and extrinsic curvature fields of the brane. Then we deduce the effective field theory on the brane in terms of those fields and the quantum fluctuations of the solitonic solution. The gravity, gauge and the extrinsic curvature fields should obey the Gauss-Codazzi-Ricci equation in addition to their own equations of motion.
\end{abstract}

\vskip15pt
]

\narrowtext
\fi

This contributed paper \footnote{
Contributed paper for ICHEP2000(XXXth International Conference on High Energy Physics
July 27 - August 2, 2000, Osaka, Japan)} 
	is based on the work \cite{1} done in collaboration 
	with Takashi Hattori.
Recently the idea of brane world attracts much attention
	in connection with the expectation 
	for ``large scale extra dimensions"
	to solve the hierarchy problems 
	between the weak interaction scale and the Planck scale, 
	and other problems \cite{ADD,DDG,Bachas,RS}.
The idea that our world is an embedded object 
	in a higher dimensional spacetime is very old.
The early works were connected with 
	embeddability problem of a Riemanian spacetime 
	in a higher dimensional spacetime \cite{early}. 
Joseph presented a picture of dynamical trapping of matters
	by a potential valley along the embedded world \cite{Joseph}.
Regge and Teitelboim pointed out that we should adopt the metric
	but not the position of the brane as the dynamical variable
	to have ordinary theory of gravity \cite{RegTei}.
Maia investigated symmetry aspects of embedded spacetimes \cite{Maia}.
Field theoretical trapping mechanisms of the world 
	were considered by the present author \cite{Nara}, 
	and Rubakov and Shaposhnikov \cite{RubShap}, independently.
The former used, as an example, the Nielsen-Olesen vortex type mechanism  
	to trap the world on the brane, 
	and the gravity is naturally described by the induced metric, 
	whose kinetic term (Einstein gravity term)
	is induced through the quantum fluctuations. 
On the other hand, the latter adopted 
	the example of the domain-wall type solution
	with special interest in the trapped bosons and chiral fermions
	on the flat brane.    
Gravitational trapping is considered by Visser \cite{Visser}.
Various models were investigated in connection 
	with various problems,
	such as chiral anomaly \cite{anom}, supermembrane \cite{supmem}, 
	and others \cite{various,Akama87}.
It turned out that the D-brane  
	play important rolls in the superstring theory \cite{Dbrane},
	or in reducing it from the M-theory \cite{Mtheory}.
Triggered by the ideas of large extra dimension \cite{ADD,DDG,Bachas,RS},
	an explosion of papers came out 
	including phenomenological \cite{pheno}, cosmological \cite{cosm}, 
	and theoretical investigations \cite{theor}. 
Besides the various possibilities as are recently studied,
	the idea of the brane world is basically important 
	because it gives an alternative to the Klein-type
	compactification \cite{Klein} 
	to hide the Kaluza-type extra dimensions \cite{Kaluza}, 
	which become necessary for various theoretical reasons.
Other important applications of the studies of physics on the brane 
	are those to the lower dimensional objects 
	such as domain walls, vortices, topological defects, etc.\ 
	in the condensed matter systems \cite{condmat}.	
Here we investigate how and what field theoretical ingredients are induced 
	on the brane world.  
We show that it gives a unified picture for gravity and gauge fields
	with additional constraints from the Gauss-Codazzi-Ricci equations.

If our world is really on a brane localized by some higher dimensional dynamics,
	the gravitation on the brane
	should be caused by dynamical deformations of the brane itself.
In addition to the gravitational fields, 
	the deformation of the brane give rise to 
	the extrinsic curvature 	and the normal connection, 
	which, respectively, behave like a tensor field 
	and a gauge field on the brane.    
This suggests a natural and challenging scenario of geometrical unification 
	of the gravitational and the gauge fields \cite{Maia,Akama87}. 
We call them unified because they are described by 
	different components of the same connection 
	from the whole spacetime viewpoints, 
	as we will explain below.

\def\bm#1{{\bf\mbox{\boldmath $#1$}}}
Let	$\bm{\omega_{\mu\nu\lambda}}$ be the affine connection,
	and $\bm{R^\mu{}_{\nu\rho\sigma}}$ be the curvature tensor of 
	the $p+q+1$-spacetime.
Then, 
they are related by 
\begin{eqnarray}&&
\bm{R_{ab\mu\nu}}=
	\bm{\omega_{ab[\mu,\nu]}}+\bm{\omega_{ca[\mu}\omega^c{}_{b\nu]}},
\label{R}
\end{eqnarray}
where the Greek suffices indicate the spacetime indices,
	and the Latin, the local Lorentz indices, 
	the square brackets [\ ] in the array of the suffices indicate
	anti-symmetrization of the suffices, i.e.
	we subtract the term with the suffix at the right of [ and 
	that at the left of ] exchanged, and 
	the comma in the array of the suffices indicates the differentiation
	with respect to the spacetime coordinate components 
	indicated by the suffices after the comma.
We use the metric $\bm{g_{\mu\nu}}$ and its inverse $\bm{g^{\mu\nu}}$
	to raise and lower spacetime indices,
	the flat metric 
	$\bm{\eta_{ab}}=\bm{\eta^{ab}}={\rm diag}(1,-1,\cdots,-1)$ 
	to raise and lower local Lorentz indices,
	and the vielbein $\bm e_{a\mu}$ 
	to convert spacetime indices to local Lorentz ones,
	and vice versa.
Here and hereafter we denote the quantities of concerning $p+q+1$-spacetime
	by boldface letters, 
	and the boldface suffices run over the range $0,1,\cdots,p+q$.

For a given $p$-brane, let us take a coordinate system $\bm x^\mu$
	such that $x^{\underline\mu}=0$ for $\underline\mu = p+1,\cdots,p+q$.
Hereafter the bold indices like $\bm \mu$, $\bm a$, etc.\ 
	run over the range $0,1,\cdots,p+q$ (whole spacetime),
	normal indices without underline like $\mu$, $a$, etc.\ 
	run over the range $0,1,\cdots,p$ (our spacetime),
	and 
	normal indices with underline like $\underline\mu$, $\underline a$, etc.\ 
	run over the range $p+1,\cdots,p+q$ (extra dimensions).
Let	${\omega_{\mu\nu\lambda}}$ be the affine connection,
	and ${R^\mu{}_{\nu\rho\sigma}}$ be the curvature tensor of 
	the $p+1$-spacetime on the $p$-brane.
Then, they are related by 
\begin{eqnarray}&&
R_{ab\mu\nu}=
	\omega_{ab[\mu,\nu]}+\omega_{ca[\mu}\omega^c{}_{b\nu]}.
	\label{Rb}
\end{eqnarray}
The brane affine connection ${\omega_{\mu\nu\lambda}}$ is nothing but the 
	brane-parallel components of the whole-spacetime connection
	$\bm{\omega_{\mu\nu\lambda}}$ at the brane.
The components $\bm\omega_{\underline a b\mu}\equiv B_{\underline a b\mu}$ 
	of the whole-spacetime connection
	with normal index $\underline a$ and tangent index $ b$ are 
	(index-converted form of) the extrinsic curvture $B_{\underline a\mu\nu}$
The components $\bm\omega_{\underline{ab}\mu}\equiv A_{\underline{ab}\mu}$ 
	of the whole-spacetime connection
	with normal indices $\underline a$ and $\underline b$ are 
	the normal connection, 
	which play the roles of the gauge fields 
	of the normal space rotation group $O(q)$. 
The curvature tensor with respect to the gauge transformation
	is the field strength of the gauge field:
\begin{equation}
{F_{\underline{ab}\mu\nu}}=
	{A_{\underline{ab} [\mu,\nu]}}
	+{A_{\underline{ca} [\mu}A^{\underline{c}}{}_{\underline{b}\nu]}}.
\end{equation}
Then (\ref{R}) implies that 
\begin{equation}
{\bm{\omega _{ ab}}_{[\mu,\nu]}}
	+{\bm{\omega _{ca}}_{[\mu}\bm{\omega ^c{}_b}_{\nu]} } 
	={\bm{R_{ab}}_{\mu\nu}}|_{x^{\underline \mu}=0},			\label{GCR}
\end{equation}
	which gives constraints 
	among the metric $g_{\mu\nu}$ 
	(or vielbein $e_{a\mu}$),
	the extrinsic curvature $B_{\underline a\mu\nu}$, 
	and the normal-connection gauge field $A_{\underline{ab}\mu}$
	in terms of the $p+q+1$ curvature at the $p$-brane.
In fact, the tangent-tangent($ab$), 
	tangent-normal($a\underline b$), 
	and normal-normal($\underline {ab}$),
	components of (\ref {GCR}) are rewritten as
\begin{eqnarray}&&
R_{ab\mu\nu}+{B_{\underline ca[\mu}B^{\underline c}{}_ {b \nu]} } 
	={\bm{R}_{ ab \mu\nu}}|_{x^{\underline \mu}=0},			\label{Gauss}
\\&&
{B_{\underline ab[\mu,\nu]}}
	+{B_{c\underline a[\mu}\omega^c{}_{b\nu]} } 
	+{A_{\underline c\underline a[\mu}B^{\underline c}{}_{b\nu]} } 
	={\bm R_{\underline ab\mu\nu}}|_{x^{\underline \mu}=0},			\label{Godazzi}
\\&&
F_{\underline{ab}\mu\nu}+{B_{ c\underline a[\mu}B^{ c}{}_ {\underline b 	\nu]} } 
	={\bm{R}_{\underline{ab} \mu\nu}}|_{x^{\underline \mu}=0},			\label{Ricci}
\end{eqnarray}
	which are called Gauss, Codazzi, and Ricci equations, respectively.
On the other hand, for the whole-spacetime with constant curvature,
	these equations are the integrability condition 
	for the solution for the $p$-brane to exist 
	uniquely up to global translations and rotations
	for a given set of the $e_{a\mu}$, $B_{\underline a\mu\nu}$,
	and $A_{\underline{ab}\mu}$.
Owing to this theorem, we can adopt the the $e_{a\mu}$, $B_{\underline a\mu\nu}$
	and $A_{\underline{ab}\mu}$ as the dynamical variables 
	of the $p$-brane system
	instead of the position $\bm y(x^\mu)$ avoiding 
	the Regge-Teitelboim problem \cite{RegTei}.

Now we consider the case where the $p+q+1$-spacetime is flat.
Let $\bm {X_\mu}$ be the Cartesian coordinate system that we refer to. 
We denote the position of the $p$-brane by $\bm y(x^\mu)$,
	where $x^\mu$ ($\mu=0,1,\cdots,p$) is $p+1$ parameters.
We define 
	$p+1$ orthonormal tangent vectors $\bm n_a(x^\mu)$ ($a=0,1,\cdots,p$) 
	$q$ orthonormal normal vectors $\bm n_{\underline a}(x^\mu)$ 
	(${\underline a}=p+1,\cdots,p+q$) of the $p$-brane.
Then the induced vielbein on the $p$-brane
	is given by 
$$e_{a\mu}=\bm n_a\bm y_{,\mu},$$
	the induced metric on the $p$-brane is given by 
$$g_{\mu\nu}=\bm y_{,\mu}\bm y_{,\nu}=e_{a\mu}e^a{}_\nu, $$
	and the induced connection on the $p$-brane is given by 
$$\omega\bm {_{ab}}{}_\mu=\bm {n_a n_b}{}_{,\mu},$$
	where the differentiation denoted by the index after the comma
	is that with respect to the parameter $x^\mu$,
	and the inner product of the two vectors 
	are taken with respect to the metric of
	the curved $p+q+1$-spacetime.

Now we transform the coordinate system from $\bm {X^\mu}$ to
	$\bm {x^\mu}$ such that
	the first $p+1$ components $x^\mu$ ($\mu=0,1,\cdots,p$)
	coincide with the parameter $x^\mu$ 
	and the last $q$ components 
	$x^{\underline \mu}$ ($\underline\mu=p+1,\cdots,p+q$)
	vanish (i.e. $x^{\underline \mu}=0$) 
	on the $p$-brane. 

We know various models which has solitonic solution 
	where fields are localized in the neighborhood of 
	some lower dimensional subspace. 
For example, we have the kink solution for a scalar field 
	in a double well potential ($q=1$),
	the Neilsen-Olesen vortex solution
	in the Higgs model with $U(1)$ gauge field ($q=2$),
	the Skirmion solution with massive scalar fields ($q=3$),
	etc..
In many cases, the flat space solution is well known.
So here we develop a general formalism
	to derive the curved brane solution in the curved spacetime
	from the well known flat-space solution.
Consider the system of the fields $\bm\Phi$
	(in general, including higher spin and spinor fields)
	with the flat-space action
\begin{equation}
S=\int L(\bm{\eta_{a\mu}},\bm\Phi,\partial\bm{_k\Phi})d_{p+q+1}\bm X,
		\label{Sflat}
\end{equation}
	which has some symmetry with non-symmetric potential minima.
Let us assume that the equation of motion of (\ref{Sflat})
	has a static soliton solution 
	$\bm \Phi=\bm\Phi_{\rm sol}(x^{\underline a})$
	such that 
\begin{equation}
\bm\Phi_{\rm sol}\sim\bm\Phi_{\rm min}+O(e^{-\delta})\ \ {\rm for}\ \ 
	r=\sqrt{-x^{\underline a} x_{\underline a }}\rightarrow \infty,
	\label{Phisol}
\end{equation}
	where $\delta $ is a constant corresponding to the soliton size, 
	and $\bm \Phi_{\rm min} $ is a 
	singular function
	which takes the value with the potential minimum 
	except at $r=0$ and is singular at $r=0$.
Around the static solution $\bm \Phi=\bm\Phi_{\rm sol}(x^{\underline a})$,
	we can find small fluctuation modes $\bm\xi$
	for which $\bm\Phi_{\rm sol}+\bm\xi$ is a solution
	of the linearized equation of motion 
	under the assumption that $\bm\xi$ is very small.
If the action has some symmetry,
	there exist zero-mode solutions associated with the symmetry.
They are still static implying zero additional energy to that of 
	$\bm\Phi_{\rm sol}$
Among them the zero-mode of the translation invariance in the extra spase
	is given by 
	$\bm\xi_{\underline a}=\partial_{\underline a}\bm\Phi_{\rm sol}$.
Other types of the small fluctuations
	are non-static ones where $\bm\xi$ depends also on $x^\mu$,
	and has additional energies.

Now we seek for the solution localized in the neighborhood of the
	given $p$-brane specified by the vielbein $e_{a\mu}$, 
	the extrinsic curvature $B_{\underline a \mu\nu}$, 
	and the normal connection gauge field $A_{\underline{ab}\mu}$.
We assume the curvature $R_{ab\mu\nu}$, 
	the extrinsic curvature $B_{\underline a \mu\nu}$, 
	and the normal connection gauge field $A_{\underline{ab}\mu}$
	are small in the scale of the soliton size $\delta$.
This is an safe and plausible assumption
	from both cosmological and field theoretical points of view.
Then the solution is obtained by the following procedure.
First we specify the curvilinear coordinate system $\bm {x^\mu}$ a little more precise
	by the transformation:
\begin{equation}
\bm X=\bm y(x^\mu)+x^{\underline a}\bm n_{\underline a}(x^\mu).
\label{X}
\end{equation}
In general the transformation (\ref{X}) becomes singular 
	at some large $|x^{\underline a}|$.
It is safe now, however, owing to the rapidly approaching property in (\ref{Phisol}) 
	and assumption of small $R_{ab\mu\nu}$, 
	$B_{\underline a \mu\nu}$, and $A_{\underline{ab}\mu}$.
Then the vielbein $\bm{e_{a\mu}}=\bm{n_a} \bm{X_{,\mu}}$ is given by
\begin{equation}
\pmatrix{
	\bm e_{a\mu} &\bm e_{a\underline\mu}\cr
	\bm e_{\underline a\mu} &\bm e_{\underline {a \mu}}
	}
	=\pmatrix{
	e_{a\mu}-x^{\underline b}B_{\underline b a\mu} &0\cr
	-x^{\underline b}A_{\underline {ba}\mu} &\eta_{\underline {a \mu}}
	}.
\label{e}
\end{equation}
The action becomes
\begin{equation}
S=\int\bm e L(\bm{e_{a\mu}},\bm\Phi, \bm{ D _k\Phi}) d_{p+q+1}\bm x 
		\label{S}
\end{equation}
	in a given curved spacetime specified by the vielbein $\bm {e_{a\mu}}$,
	where $\bm e =\det\bm {e_{a\mu}}$, and $ \bm{ D _{\mu}}$ is the covariant 
	differentiation written in terms of the connection $\bm{\omega_{ab\mu}}$
	according to the spins of the ingredients of the field $\bm \Phi$.

If the $p$-brane is curved, the solution $\bm\Phi_{\rm sol}$
	is no longer a solution of the equation of motion.
The small correction is expected to have an asymptotic form
	proportional to the translation zero-mode.
Then we parametrize the small fluctuation from $\bm\Phi_{\rm sol}$ 
	as
\begin{equation}
\bm\Phi_{\rm brane}(x\bm{^\mu})
	=\bm\Phi_{\rm sol}(x^{\underline \mu})
	+\bm\chi^{\underline \mu}\partial_{\underline \mu}
	\bm\Phi_{\rm sol}(x^{\underline \mu}),
\label{Phibrane}
\end{equation}
	where $\bm\chi^{\underline \mu}$ is the small coefficient 
	(in general matrix because $\bm\Phi$ is a system of fields)
	which depends on both $x^\mu$ and $x^{\underline \mu}$.
Because the solution rapidly approaches its asymptotic value for
	$x^{\underline \mu}>>\delta$, 
	only the region with small $x^{\underline \mu}$
	is important.
So we expand $\bm\chi$ in terms of $ x^{\underline \mu}$:
\begin{equation}
\bm\chi=
	\bm\chi^{(0)}+x^{\underline \mu}\bm\chi^{(1)}_{\underline \mu}
	+ x^{\underline \mu} x^{\underline \nu}\bm\chi^{(2)}_{\underline {\mu\nu}}
	+\cdots,	\label{chi}
\end{equation}
	where $\bm\chi^{(i)}$ ($i=0,1,2,\cdots$) are functions of $x^\mu$ only.
The function $\bm\chi^{(0)}$ corresponds to the shift of the position
	of the brane, and it should be described by
	other values of $e_{a\mu}$, 
	$B_{\underline a\mu\nu}$, and $A_{\underline{ab}\mu}$.
Therefore we neglect $\bm\chi^{(0)}$ in (\ref{chi}).
Now we can expand the equation of motion 
	in power series of $x^{\underline \mu}$.
The equation should hold order by order in $x^{\underline \mu}$.
This gives recursive equations for the functions $\bm\chi^{(i)}$
	in terms of the vielbein $e_{a\mu}$, 
	the extrinsic curvature $B_{\underline a \mu\nu}$, 
	and the normal connection gauge field $A_{\underline{ab}\mu}$.
Thus we can obtain the $p$-brane solution $\bm\Phi_{\rm brane}(x\bm{^\mu})$
	in the form of (\ref{Phibrane}),
	where $\bm\chi$ is power series of $x^{\underline \mu}$
	with coefficients written in terms of the parameters
	$e_{a\mu}$, $B_{\underline a \mu\nu}$, 
	and $A_{\underline{ab}\mu}$ which specify the $p$-brane. 
If we substitute this $\bm\Phi_{\rm brane}(x\bm{^\mu})$ 
	back into the action (\ref{S})
	and perform $x^{\underline\mu}$-integration,
	we obtain the effective action for the quantities
	$e_{a\mu}$, $B_{\underline a \mu\nu}$, 
	and $A_{\underline{ab}\mu}$ on the brane world.

As an example, here we apply this procedure to the model
	for the scalar field $\Phi$ 
	with a double well potential
	in the five dimensional spacetime \cite{RubShap}.
The Lagrangian is given by
\begin{equation}
L=\bm e\left[\frac{1}{2}\bm{g^{\mu\nu}}
\partial\bm{_\mu}\Phi\partial\bm{_\nu}\Phi
-\frac{1}{4}\lambda\left(\Phi^2-\frac{m^2}{\lambda}\right)^2\right],
\end{equation}
where 
$\lambda$ is the coupling constant and $m$ is the mass of the field $\Phi$.
In the flat case, this is known to have the domain wall solution 
\begin{equation}
\Phi=\Phi_{\rm sol}\equiv\frac{m}{\sqrt{\lambda}}\tanh\frac{my}{\sqrt2},
\end{equation}
where we have denoted the extra dimension coordinate $x^{\underline 4}$ by $y$.
Then, after a lengthy calculation according to the above procedure, 
	we obtain the brane world solution
\begin{eqnarray}&&
\Phi_{\rm brane}=\Phi_{\rm sol}
\cr&&
+\left[\frac{1}{2}y^2 B_\mu^\mu +\frac{1}{6}y^3
\left((B_\mu^\mu)^2+B_{\mu\nu} B^{\mu\nu}
\right) +O(B^3)
\right]
\Phi'_{\rm sol},
\end{eqnarray}
where 
$B_{\mu\nu}$ is the extrinsic curvature, and
$\Phi'_{\rm sol}$ is the derivative of $\Phi_{\rm sol}$
with respect to the extra dimension coordinate $y$.

Now we return to the general formalism.
Let $\bm\xi_n$ ($n=1,2,\cdots$) be the small fluctuation modes 
	of the flat space solution.
We expand the small fluctuation of
	the field $\Phi$ in terms of $\bm\xi_n$ as
\begin{equation}
\bm\Phi=
	\bm\Phi_{\rm brane}
	+\sum_n\bm\varphi_n(x^\mu)\bm\xi_n(x^{\underline \mu}).
\end{equation}
We substitute this into the action (\ref{S}),
	and perform the $x^{\underline\mu}$-integration.
Then we obtain the effective action $S_{\rm eff}$ 
	for $e_{a\mu}$, $B_{\underline a \mu\nu}$, 
	$A_{\underline{ab}\mu}$, and $\bm\varphi_n$
	on the $p$-brane.

So far we have treated the fields $\bm{e_{a\mu}}$ in the $p+q+1$-spacetime
	and $e_{a\mu}$, $B_{\underline a \mu\nu}$, 
	and $A_{\underline{ab}\mu}$ on the brane
	as the given external fields.
To incorporate their dynamics, we should path-integrate 
	the generating functional 
	over all the possible configurations.
We can add the kinetic term of $\bm{e_{a\mu}}$ in the $p+q+1$-spacetime
	because it is taken as an elementary field.
Then we will have some equation which specifies 
 	the configuration of $\bm{e_{a\mu}}$ in the $p+q+1$-spacetime.
If the quanta (gravitons etc.) of the system is massless,
	the coupling of the interaction should not be so strong 
	as to violate the observed level of the effective energy conservation 
	on the brane.
On the other hand, we should path-integrate the generating functional
 	with respect to $e_{a\mu}$, $B_{\underline a \mu\nu}$, 
	and $A_{\underline{ab}\mu}$ on the brane 
	to incorporate all the possible and independent configurations
	of the brane.
However, the effective action derived as above has no kinetic term
	of $e_{a\mu}$, $B_{\underline a \mu\nu}$, 
	and $A_{\underline{ab}\mu}$.
The kinetic terms are induced through the quantum fluctuations 
\cite{Sakharov,ACMT,Akama78,Adler}
	of the trapped matter field $\bm\varphi$ on the brane.
In general the quantum fluctuations have the form 
\begin{equation}
\ e(a + b R+\cdots), 
\end{equation}
where $a$ and $b$ are calculable constants.
The field theoretical calculations are found in the literature 
	\cite{ACMT,Akama78,Adler}.
The values of $a$ and $b$, respectively, 
	contain quartic and quadratic divergences in the ultraviolet region, 
	which we cutoff at around the scale 
 	of asymptotic field value $|\bm\Phi_{\rm min}|$. 
Such a cutoff is naturally expected in the present model, 
	since very high excitation above the threshold of $|\bm\Phi_{\rm min}|$
	spread into the extra dimensions and blind to the position of the brane.
Then the effective action on the brane gives the equations of motion
 	of $e_{a\mu}$, $B_{\underline a \mu\nu}$, 
	and $A_{\underline{ab}\mu}$ on the brane.
The cosmological constant should be fine-tuned.

In conclusion, on the brane world localized by the solitonic solution,
\\(i) the small fluctuation modes trapped by solitonic solution 
play the role of matter fields on the brane,
\\(ii) the gravitational and the gauge fields are induced 
through the deformation of the brane,
\\(iii) besides them the extrinsic curvature field are induced 
through the deformation of the brane,
\\(iv) the kinetic terms of the gravitational, the gauge and
the extrinsic curvature fields are induced 
through the quantum fluctuations of the trapped matter fields, and
\\(v) the gravitational, the gauge and
the extrinsic curvature fields should obey the 
Gauss-Codazzi-Ricci equations in addition
to the ordinary equations of motion.

\end{document}